\documentclass[epj]{svjour}
\usepackage{bm}
\usepackage{epsfig,graphics}
\usepackage{amsfonts,amssymb,stmaryrd,latexsym,amsmath}
\usepackage{slashed}

\newcommand{\be}{\begin{equation}}
\newcommand{\ee}{\end{equation}}
\newcommand{\ba}{\begin{eqnarray}}
\newcommand{\ea}{\end{eqnarray}}

\begin{document}

\title{The size of the proton - closing in on the radius puzzle}
\author{I.~T.~Lorenz\inst{1}\thanks{{\it Email address:} lorenzi@hiskp.uni-bonn.de},
        H.-W.~Hammer\inst{1}\thanks{{\it Email address:} hammer@hiskp.uni-bonn.de},
        and Ulf-G. Mei{\ss}ner\inst{1,2}\thanks{{\it Email address:} meissner@hiskp.uni-bonn.de}}
\institute{Helmholtz-Institut f\"ur Strahlen- und Kernphysik and
          Bethe Center for Theoretical Physics, Universit\"at Bonn, D-53115
          Bonn, Germany
          \and
          Institute for Advanced Simulation and J\"ulich Center for Hadron
          Physics, Institut f\"{u}r Kernphysik, Forschungszentrum J\"{u}lich, D-52425 J\"{u}lich,
          Germany}

\date{\today}

\abstract{We analyze the recent electron-proton scattering data from Mainz
using a dispersive framework that respects the constraints from
analyticity and unitarity on the nucleon structure. We also perform a
continued fraction analysis of these data. We find a small electric proton
charge radius,  $r_E^p = 0.84_{-0.01}^{+0.01}~{\rm fm}$, consistent with the
recent determination from muonic hydrogen measurements and earlier dispersive
analyses. We also extract the proton magnetic radius,
$r_M^p = 0.86 _{-0.03}^{+0.02}~{\rm fm}$, consistent with earlier
determinations based on dispersion relations.}


\authorrunning{I.~T.~Lorenz {\it et al.}}
\titlerunning{The size of the proton}

\maketitle

\medskip

The proton charge radius is a  fundamental quantity of physics. It is truly
remarkable that despite decade long experimental and theoretical efforts its
precise value is not yet determined. The recent controversy about the size
of the proton was triggered by the precision measurement of the Lamb shift
in muonic hydrogen that led to a ``small''  charge radius, $r_E^p =
0.84184(67)\,$fm \cite{Pohl:2010zza}. This result came as a big surprise as it
was in stark contrast to the commonly accepted ``large'' CODATA  value of 
$0.8768(69)\,$fm \cite{Mohr:2008fa}, based on measurements of the Lamb shift
in  electronic hydrogen and the analysis of electron-proton scattering data.
The large value was further strengthened by the high precision electron-proton
scattering measurements at MAMI-C \cite{Bernauer:2010wm,Bernauer:2011zza}.
The analysis of these data including two-photon corrections led to   
$r_E^p=0.876(8)\,$fm.
These authors also found a magnetic radius of the proton that came out much 
smaller than commonly accepted values, 
$r_M^p=0.803(17)\,$fm.

On the theoretical side, a precise ab initio calculation based on lattice
QCD is not yet available due to various conceptual problems to be overcome
like the treatment of disconnected contributions that feature importantly
in isoscalar quantities. See, e.g.,~\cite{Collins:2011mk} for a recent work. A
different framework, that was pioneered by H\"ohler and collaborators a long
time ago, are dispersion relations (DRs) for the nucleon electromagnetic form
factors \cite{Hohler:1974eq}. In such type of approach, smaller radii were
always favored, and the most recent and sophisticated DR calculation led
to values consistent with the ones from muonic hydrogen, $r_E^p =
0.844_{-0.004}^{+0.008}\,$fm \cite{Belushkin:2006qa}. 
The value for
the proton's magnetic radius was $r_M^p= 0.854\pm 0.005\,$fm, consistent
with, e.g., the determination based on continued fractions by Sick
\cite{Sick:2005az}.
Note, however, that the conformal mapping technique to analyze the nucleon form factors
led to a large value of the proton charge radius~\cite{Hill:2010yb}.

Given this puzzling situation, in this Letter we will re-analyze the MAMI 
data of Bernauer et al. \cite{Bernauer:2010wm}
using dispersion relations. Our main focus will be
on the correct treatment of the analytical structure of the nucleon form
factors that is driven by the two-pion continuum. Its important role was
already stressed by Frazer and Fulco, who were able to predict the
$\rho$-resonance and its influence on the nucleons' structure a long time ago
\cite{Frazer:1959gy}. What has often been overlooked since this seminal
work was the large enhancement of the two-pion continuum on the left wing
of the $\rho$-resonance due to a close-by pole on the second Riemann sheet in the elastic
pion-nucleon scattering amplitude. This enhancement amounts for roughly half of the
nucleon isovector size   \cite{Hohler:1974eq}. 
This important effect is also recovered in chiral
perturbation theory, the effective field theory of QCD at low energies
\cite{Bernard:1996cc}. We consider it therefore of utmost 
importance to include this effect in the re-analysis of the MAMI data.
Of course, in light of the muon $g-2$ measurement at BNL \cite{Bennett:2006fi}, 
one might speculate about the influence of new physics as the muon data are more sensitive 
to such effects, but first one has to exclude possible conventional
explanations -- and we will offer such a possibility here.

\vfill
\newpage

The nucleon  matrix element of the electromagnetic current 
\begin{eqnarray}\label{eq:em_curr}
&&\!\!\!\!\!\!\!\!\!\!\!\!\!\!\!\!\!\!\!\! 
\langle N(p')|{j}_{\mu}^{\rm em}|N(p)\rangle \nonumber\\
&& = ie\bar{u}(p')\biggl(\gamma_{\mu}F_1(t)+i\frac{\sigma_{\mu\nu}q^{\nu}}{2m_N}F_2(t)\biggr)u(p),
\end{eqnarray}
is parameterized in terms of the Dirac, $F_1(t)$, and Pauli, $F_2(t)$, form
factors, with $t = (p'-p)^2 = -Q^2$ the invariant momentum transfer squared,
and $m_N$ is the nucleon mass. For electron-nucleon scattering, we have
$Q^2\geq 0$. $F_1^{p/n}(0)$ and $F_2^{p/n}(0)$ are given in terms of the 
proton/neutron electric charge and anomalous magnetic moment, respectively.
For the later analysis, it is useful to separate the form factors in their
isoscalar and the isovector parts,
$ F_i^{s} = (F_i^p+F_i^n)/2$ and $F_i^{v} = (F_i^p-F_i^n)/2$ for $i =1,2$,
correspondingly. The differential cross section is given most compactly
in terms of the Sachs form factors $G_E(t) = F_1(t)-\tau F_2(t)$, 
$G_M(t) = F_1(t)+F_2(t)$, with $\tau = -t/4m_N^2$, so that
\begin{equation}\label{eq:xs_ros}
\frac{d\sigma}{d\Omega} = \left( \frac{d\sigma}
{d\Omega}\right)_{\mathrm Mott} \frac{\tau}{\epsilon (1+\tau)}
\left[G_{M}^{2}(Q^{2}) + \frac{\epsilon}{\tau} G_{E}^{2}(Q^{2})\right]\, ,
\end{equation}
where $\epsilon = [1+2(1+\tau)\tan^{2} (\Theta/2)]^{-1}$ is the virtual photon polarization,
$\Theta$ is the electron scattering angle in the
laboratory frame, and $({d\sigma}/{d\Omega})_{\mathrm Mott}$ is the Mott cross section, which
corresponds to scattering on a point-like particle.
The electric and magnetic radii of the proton, that are in the focus of
this Letter, are given by
\begin{equation}\label{eq:rad}
r_{E,M}^p = \left(\left.\frac{-6}{G_{E,M}(0)}\frac{dG_{E,M}(Q^2)}{dQ^2}\right|_{Q^2 = 0}\right)^{1/2}~.
\end{equation}

To analyze the cross section data from Mainz based
on Eq.~(\ref{eq:xs_ros}), we use  unsubtracted dispersion relations for
the nucleon form factors. For a generic form factor in the spacelike region, these take the form 
\begin{equation}
 F(t) = \frac{1}{\pi}\int_{t_0}^{\infty}\frac{\text{Im}F(t')dt'}{t'-t}~,
\label{disprel}
\end{equation}
with $t_0 = 4M_\pi^2 \, (9M_\pi^2)$ the pertinent isovector (isoscalar)
threshold and $M_\pi$ is the charged pion mass. The basic quantity in this
relation is the spectral function Im~$F(t)$ that parameterizes all 
physical effects that contribute to the nucleon form factors.
The most general form of the  spectral function allowed by unitarity 
is a sum of continua and poles. This is exactly the form of the spectral 
function we use in our analysis. The low mass continua $(2\pi,K\bar K,\rho\pi)$
are included exactly, whereas higher mass continua are approximated by
effective vector meson poles.

In our DR approach,
the complete isoscalar and isovector parts of the 
Dirac and Pauli form factors, respectively, are parameterized as
\begin{eqnarray}
 F_i^s (t) &=& \sum_{V=K\bar{K},\rho\pi,s_1,s_2,..}
\frac{a_i^V}{m_V^2-t}~,\nonumber\\
 F_i^v (t) &=& \sum_{V=v_1,v_2,..}\frac{a_i^V}{m_V^2-t} +
 \frac{a_i+b_i(1-t/c_i)^{-2}}{2(1-t/d_i)}~,\label{VMDspec}
\end{eqnarray}
where $i = 1,2$.
Each pole term comes, in principle, with  three parameters -- the two
residua and the mass. 
While the low-mass pole terms can be interpreted as physical vector mesons,
i.e. the $\omega$ and the $\phi$, the higher mass poles are effective poles that 
parameterize unknown continuum contributions. The residua for the light 
vector mesons $\omega, \phi$ are related to more or less known coupling constants,
see e.g. Ref.~\cite{Belushkin:2006qa} for a detailed discussion.
Therefore, we constrain  the residua of the light isoscalar vector mesons as:
$0.5\,\mbox{GeV}^2<|a_{1}^\omega| < 1\,\mbox{GeV}^2$, $|a_{2}^\omega| < 
0.5\,\mbox{GeV}^2$~\cite{Grein:1977mn}
and $|a_{1}^\phi| < 2\,\mbox{GeV}^2$, $|a_{2}^\phi| < 1\,\mbox{GeV}^2$~\cite{Meissner:1997qt}.
Additional poles below $~1\,\mbox{GeV}^2$ can be excluded since no other vector mesons
exist in this region. At higher masses, the widths get broader and with growing distance to
the space-like region their position cannot be resolved precisely. Therefore this part of
the spectral function can only be described by effective pole terms.
Besides the vector meson pole terms, our DR approach 
includes the two-pion
continuum, the $K\bar{K}$- and $\rho\pi$-continua. The two latter
contributions to the form factors are also parameterized by pole terms with fixed masses
and residua, as was done before, see
Refs.~\cite{Hammer:1999uf,Hammer:1998rz,Meissner:1997qt}.
The last term in 
Eq.~\eqref{VMDspec} is an appropriate parameterization of the two-pion
continuum, that also includes the $\rho$-pole as discussed above. The
parameters are: 
\begin{eqnarray*}
a_1 &=& 1.084, a_2 =5.800, b_1= 0.079, b_2 = 0.751,\\ 
c_1 &=& 0.300\,{\rm GeV}^2, c_2 = 0.225\,{\rm GeV}^2,\\ 
d_1 &=& 0.522\,{\rm GeV}^2, d_2 = 0.562\,{\rm GeV}^2~. 
\end{eqnarray*}
Note that this parameterized
form of the low-mass continua is only used for faster numerical evaluations.
The total number
of isoscalar and isovector poles is determined from the stability criterion of
Ref.~\cite{Sabba Stefanescu:1978sy}, which in a nutshell can be described as
using the lowest numbers of parameters that are required to obtain a good fit
to the data. The superconvergence relations enforcing the correct 
asymptotic power behavior of the form factors,
\begin{equation}
\int_{t_0}^\infty {\rm Im}\,F_i (t)\, t^n\,dt =0\,,\quad i=1,2,
\label{eq:sc}
\end{equation}
with $n=0$ for $F_1$ and $n=0,1$ for $F_2$, are included as well.

With this DR ansatz, we reconstruct the differential cross
sections and fit to the MAMI data for $ep$ scattering and simultaneously to
the world data for the neutron form factors. The fits are done in the
one-photon approximation via the Rosenbluth formula where the data contains 
most of the higher order corrections \cite{Bernauer:2010wm}. However, the 
Coulomb corrections are replaced in our analysis by a better
approximation  as was done by \cite{Sick:2003gm}. For a recent discussion on
this point, see Refs.~\cite{Arrington:2011kv,Bernauer:2011zza}. We remark that
two-photon exchange corrections are expected to be small for the charge radius
extraction, see eg. Refs.~\cite{Blunden:2005jv,Belushkin:2007zv},
but somewhat more sizeable for the magnetic radius 
\cite{Bernauer:2011zza,Arrington:2012dq}. This requires further study.

\begin{figure}[t]
    \centering
    \begin{tabular}{cc}
      \resizebox{41mm}{!}{\includegraphics[angle=270]{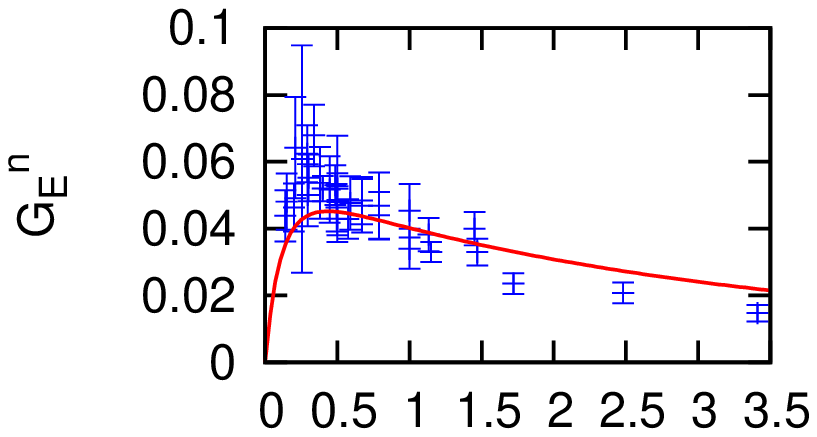}} &
      \resizebox{41mm}{!}{\includegraphics[angle=270]{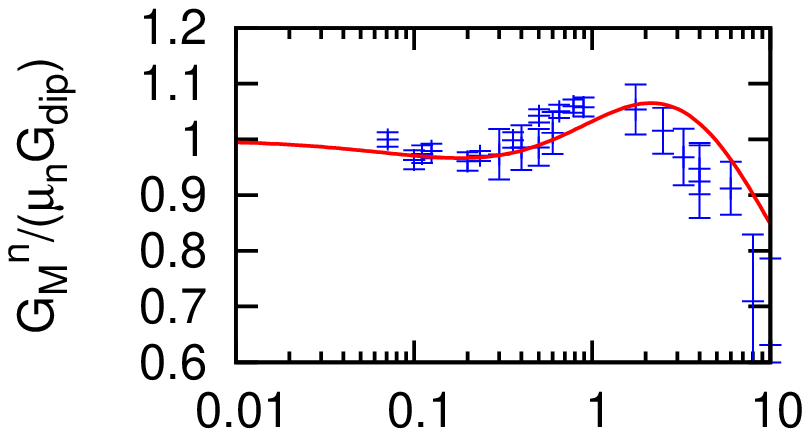}} \\
      \resizebox{41mm}{!}{\includegraphics[angle=270]{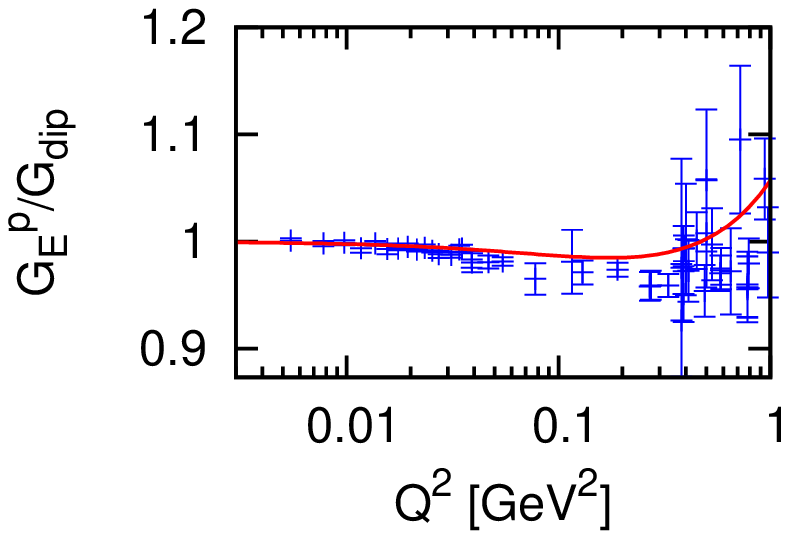}} &
      \resizebox{41mm}{!}{\includegraphics[angle=270]{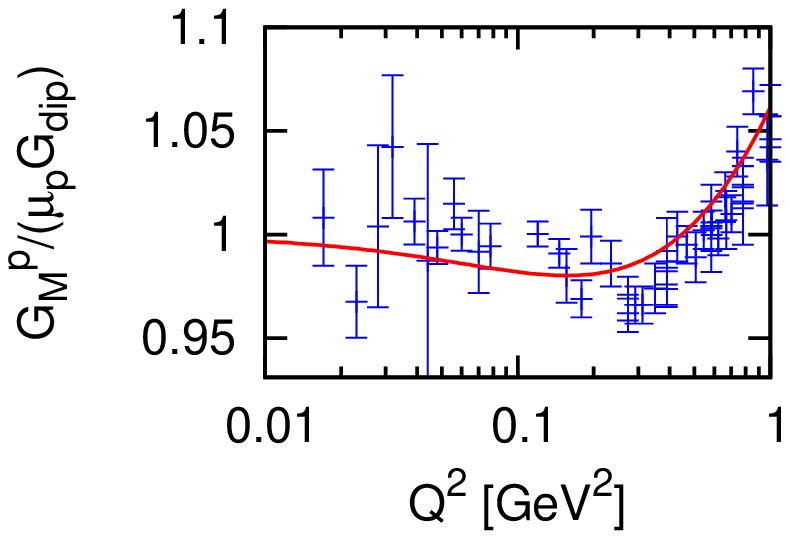}} \\
    \end{tabular}
    \vglue-2mm
    \caption{Form factors from our dispersion relation analysis in comparison
             to world form factor data as given by \cite{Belushkin:2006qa}
             (updated to include also the more recent data \cite{Riordan:2011}).\label{fig:ffs}}
    \vspace{-0.2cm}
    \end{figure}

We fit to the Mainz data floating the
normalization of the individual data sets by at most $\pm 4\%$ (in most cases
this normalization change is less than 1\%),
using the 31 normalization parameters given in Ref.~\cite{Bernauer10}.
Employing the 
aforementioned stability criterion, we include 5 isoscalar and 4 isovector 
resonances with in total 15 additional parameters. We impose the constraints
from the normalizations, from the superconvergence relations and the fact 
that the masses of the lowest isoscalar poles, the $\omega(782)$ and the 
$\phi(1020)$ are known.
For details of the fitting procedure, see Ref.~\cite{Belushkin:2006qa}.  
This fit gives a good description of the proton cross section and neutron form factor 
data with $\chi^2_{\rm red}=2.2$.
The form factors arising from this fit are shown in Fig.~\ref{fig:ffs}.
The corresponding fit parameters are given in Tab.~\ref{tab:ffs}.
    \begin{table}[ht!]
    \centering
    \begin{tabular}{|l|l|l|l||l|l|l|l|}
    \hline
    $V$ & $m_v$ & $a_1^V$ &$a_2^V$ & $V$ & $m_V$ & $a_1^V$ & $a_2^V$  \\
    \hline
    $\omega$ &        0.783        & 0.747         & 0.003      & $v_1$ &
1.205 & 0.826 & -2.557\\
    $\phi$   &        1.019        & 0.221         & -0.014     & $v_2$ &
2.140 & -0.502 & 0.438\\
    $s_1$    &        1.638        & 0.977         & 0.234      & $v_3$ &
1.000 & -0.381 & -0.342\\
    $s_2$    &        2.400        & -0.501        & -0.077     &
$v_4$ & 1.193 & -0.227 & 0.831 \\
    $s_3$    &        1.033        & -0.540        & -0.333     & & & & \\
    \hline
    \end{tabular}
    \caption{Fit parameters from dispersion analysis
             (cf.~Eq.~(\ref{VMDspec})). Masses $m_V$ are
             given in GeV and couplings $a_i^V$ in GeV$^{2}$.}\label{tab:ffs}
    \label{table:fitC}
    \vspace{-0.2cm}
    \end{table}
The obtained residua of $\omega$ and $\phi$ are in agreement with their 
couplings to the nucleon, as described above. All remaining poles have 
residua with natural size. Varying the number of pole terms in both isospin 
channels does not affect the radii significantly but is included in
our error estimate. 

To address the issue of the theoretical uncertainties of our analysis,
we vary the isovector and isoscalar continuum contributions in all possible combinations and 
repeat the fit to obtain the corresponding ranges. The two-pion continuum 
is varied by 5$\%$ below the first minimum of the pion-nucleon  scattering 
amplitudes and 20$\%$ above it. We note that at present a Roy-Steiner 
machinery is set up to improve the representation of the $t$-channel 
pion-nucleon phase shifts required here \cite{Ditsche:2012fv}, but
for our investigation we use an updated version of the two-pion
continuum representation from Ref.~\cite{Belushkin:2005ds}, including
new data for the pion vector form factor from KLOE~\cite{Ambrosino:2010bv} 
and BABAR~\cite{Aubert:2009ad}. 
The error in the two-pion continuum from the difference between the KLOE and BABAR
data sets is well below the assigned 5-20\% uncertainty.  It is also important to 
note that the Roy-Steiner analysis of Ref.~\cite{Ditsche:2012fv} has
shown that the KH80 solution used here is internally consistent.
The $K\bar{K}$- and $\rho\pi$-continua are varied by 20$\%$.
We have also varied the number of effective pole terms and
obtained stable results.

The extracted proton radii are: 
\begin{equation}\label{eq:radii}
r_E^p = 0.84_{-0.01}^{+0.01}~{\rm fm}\,,\qquad
r_M^p = 0.86 _{-0.03}^{+0.02}~{\rm fm}\,.                      
\end{equation}
For the neutron magnetic radius, we obtain $r_M^n = 0.88\pm 0.05~{\rm fm}\,,$
consistent with previous analyses. The neutron electric radius is very insensitive
to the variation of the continua. We obtain the value $(r_E^n)^2 = -0.127~{\rm fm}^2\,$.
To examine the influence of the neutron data, we also fit the DR model
exclusively to the electron-proton scattering cross sections. The effect
on the proton radii is less than half a percent on the numbers given above.
If the superconvergence relations, Eq.~(\ref{eq:sc}), are not enforced, the same
radii are obtained within the quoted uncertainties.

\begin{figure}[t]
\centering
\includegraphics[width=0.35\textwidth,angle=270]{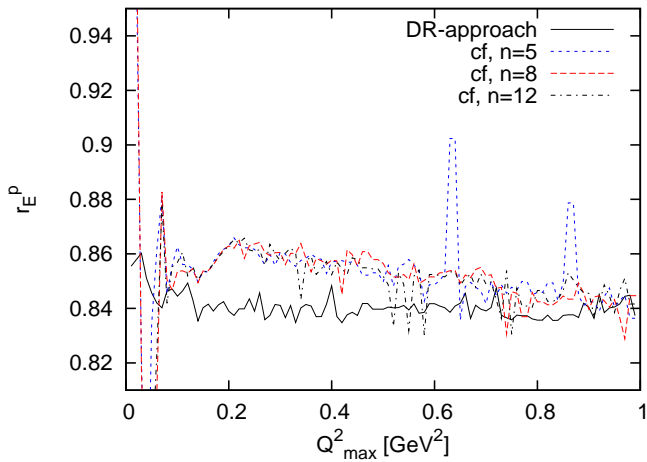} 
\caption{Dependence of the proton electric radius $r_E^p$ 
on $Q^2_{\rm max}$ for CF fits.
Here, $n$ labels the number of terms. For comparison, the DR approach 
is also shown.
\label{fig:CLrecf}}
\vspace{-0.2cm}
\end{figure}

To assess the stability of the results obtained, we make use of a continued
fraction (CF) approach as advocated, e.g., by Sick~\cite{Sick:2003gm}.
Every rational function can be represented by a general continued fraction
function and in contrast to a polynomial, this can generate poles. Therefore
it can well approximate the data beyond the threshold where the imaginary part 
sets in. We fit the following inverse CF ansatz~\cite{Sick:2003gm}, 
\begin{equation}
  F(t) =  \cfrac{1}{1
          + \cfrac{f_1t}{1
          + \cfrac{f_2t}{1 + ...}}}
\end{equation}
to our data, in order to examine under which circumstances and how exactly the
generation of isolated poles suffices to simulate the so important two-pion continuum
contribution. In particular, we have investigated the dependence of the fits
on the highest value of $Q^2$ (denoted $Q^2_{\rm max}$ from here on) for which 
data is included. This dependence can be used to demonstrate the need to
include imaginary parts in the fit functions and include data beyond the
two-pion threshold to obtain a stable fit. The electric radius derived from such fits
shows only small fluctuations for varying $Q^2_{\rm max}$, as shown in
Fig.~\ref{fig:CLrecf}. For the magnetic radius, we do not find stable results
with varying number of continued fractions and thus will not consider it any more.
Remarkably, the electric radius of the proton
show large fluctuations in the CF approach for $Q^2_{\rm max} \leq
0.1\,$GeV$^2$, but then converges quickly towards values consistent
with the ones based on the theoretically preferable DR approach. 
These results might at first appear surprising, as naively  a low
$Q^2_{\rm max}$ might be favored, because the relevant information for the radius 
extraction is related to the region of small momentum transfer. On the other
hand,  the uncertainty due to the normalization has obviously a stronger influence on
the radius for lower $Q^2$.  The less stable values at very low $Q^2_{\rm
  max}$ in the DR approach might be explained by this normalization
uncertainty  or simply by an insufficient amount of data. At higher values 
of $Q^2_{\rm max}$ the fluctuations in the derived radii increase. 
We assume that this is due to an insufficient number of free fit parameters 
for more data since the fluctuations are stronger for lower $n$.
Increasing the data range also increases the $\chi^2_{\rm red}$ slightly. Of
course, this could either show the inability of the models to describe the
data at higher momentum transfers or it could be due to a bad estimation of 
errors in this data. We cannot separate these effects, which may both apply
here: (i) it is obvious that the fit functions are only approximate and (ii)
the errors of the data are scaled to yield a $\chi^2_{\rm red} \approx 1$ for a spline
fit and thus contain a model-dependence \cite{Bernauer10}. Therefore, the
$\chi^2$ values or confidence limits are not used to weight the extracted
radii. The results are weighted equally and only the spikes of the fits with
$\chi^2_{\rm red} \simeq 1.7$ are omitted, less than 3 for each set of results. The
remaining $\chi^2_{\rm red}$ values for all fits with $Q^2_{\rm max} \geq 0.1$~GeV$^2$
vary, e.g., for the CFs of order 5 between $\chi^2_{\rm red} = 1.03$ and 1.64. The
average and standard deviation of the radius values extracted for $Q^2_{\rm
  max}$ between 0.1~GeV$^2$ and the complete data set range up to 0.98~GeV$^2$ 
are $r_E^p = 0.85 \pm 0.01\,$fm for $n=5$, $n=8$ and $n= 12$ (to obtain these results, 
we have utilized the improved Coulomb corrections as described before).
These numbers are in good agreement with the values obtained via the 
DR approach and the recent muonic hydrogen measurements. We are presently 
studying the application of the conformal mapping technique of Hill and Paz
\cite{Hill:2010yb} to the MAMI data to further sharpen our
conclusion on the value of the proton charge radius \cite{prep}.

In this Letter, we have reconsidered the proton radii
determination from the new MAMI electron-proton scattering data, using two
very different methods. First, we have used a dispersive representation of 
the nucleon form factors that includes the so important constraints from
unitarity and analyticity --  these were not considered in the fit functions
utilized in Refs.~\cite{Bernauer:2010wm,Bernauer:2011zza}. 
Including also data from the world
data set on the neutron form factors, we achieve a good description of the 
cross section data. We have also performed an analysis of the systematic
uncertainties by allowing for generous variations in the $\pi\pi$, $\bar KK$
and $\pi\rho$ contributions to the spectral functions. The resulting electric 
and magnetic radii, cf. Eq.~(\ref{eq:radii}) are in stark contrast to the ones
obtained in  Refs.~\cite{Bernauer:2010wm,Bernauer:2011zza}, but our results are consistent
with all earlier dispersive analyses of the nucleon 
form factors \cite{Hohler:1976ax,Mergell:1995bf,Hammer:2003ai,Belushkin:2006qa}. 
Also, the small value for $r_E^p$ is consistent with the recent muonic hydrogen measurement
\cite{Pohl:2010zza}. We have also applied the continuous fraction method to
the data. By construction, it can emulate narrow poles but to account for the
so important two-pion continuum, one has to choose a sufficiently large fit
range, $Q^2_{\rm max} \geq 0.1\,$GeV$^2$,
to achieve stable results. Again, this method leads to  values for the proton 
charge radius in agreement with the dispersive approach.  We conclude that the small proton
electric radius is indeed favored if analyticity and unitarity are properly
included into the description of the nucleon electromagnetic form factors.

\subsection*{Acknowledgments}
We thank Michael Distler and Ingo Sick for discussions and Achim Denig
for providing the pion form factor data. This work is  supported
in part by the DFG and the NSFC through
funds provided to the Sino-German CRC 110 ``Symmetries and
the Emergence of Structure in QCD'',
and the EU I3HP ``Study of Strongly
Interacting Matter'' under the Seventh Framework Program of the EU, and
by BMBF (Grant No. 06BN7008).


\begin{thebibliography}{99}

\bibitem{Pohl:2010zza} 
  R.~Pohl {\it et al.},
  Nature {\bf 466}, 213 (2010).

\bibitem{Mohr:2008fa} 
  P.~J.~Mohr, B.~N.~Taylor and D.~B.~Newell,
  Rev.\ Mod.\ Phys.\  {\bf 80}, 633 (2008)
  [arXiv:0801.0028 [physics.atom-ph]].

\bibitem{Bernauer:2010wm} 
  J.~C.~Bernauer {\it et al.}  [A1 Collaboration],
  Phys.\ Rev.\ Lett.\  {\bf 105}, 242001 (2010)
  [arXiv:1007.5076 [nucl-ex]].

\bibitem{Bernauer:2011zza} 
  J.~C.~Bernauer  {\it et al.},
  Phys.\ Rev.\ Lett.\  {\bf 107}, 119102 (2011).


\bibitem{Collins:2011mk}
  S.~Collins {\it et al.},
  Phys.\ Rev.\ D {\bf 84}, 074507 (2011)
  [arXiv:1106.3580 [hep-lat]].

\bibitem{Hohler:1974eq} 
  G.~H\"ohler and E.~Pietarinen,
  Phys.\ Lett.\ B {\bf 53}, 471 (1975).


\bibitem{Belushkin:2006qa} 
  M.~A.~Belushkin, H.-W.~Hammer and U.-G.~Mei\ss ner,
  Phys.\ Rev.\ C {\bf 75}, 035202 (2007)
  [hep-ph/0608337].

\bibitem{Sick:2005az} 
  I.~Sick,
  Prog.\ Part.\ Nucl.\ Phys.\  {\bf 55}, 440 (2005).

\bibitem{Hill:2010yb} 
  R.~J.~Hill and G.~Paz,
  Phys.\ Rev.\ D {\bf 82}, 113005 (2010)
  [arXiv:1008.4619 [hep-ph]].

\bibitem{Frazer:1959gy} 
  W.~R.~Frazer and J.~R.~Fulco,
  Phys.\ Rev.\ Lett.\  {\bf 2}, 365 (1959).

\bibitem{Bernard:1996cc} 
  V.~Bernard, N.~Kaiser and U.-G.~Mei\ss ner,
  Nucl.\ Phys.\ A {\bf 611}, 429 (1996)
  [hep-ph/9607428].

\bibitem{Bennett:2006fi} 
  G.~W.~Bennett {\it et al.}  [Muon G-2 Collaboration],
  Phys.\ Rev.\ D {\bf 73}, 072003 (2006)
  [hep-ex/0602035].

\bibitem{Grein:1977mn} 
  W.~Grein and P.~Kroll,
  Nucl.\ Phys.\ B {\bf 137}, 173 (1978).

\bibitem{Meissner:1997qt} 
  U.-G.~Mei\ss ner, V.~Mull, J.~Speth and J.~W.~van Orden,
  Phys.\ Lett.\ B {\bf 408}, 381 (1997)
  [hep-ph/9701296].


\bibitem{Hammer:1999uf} 
  H.-W.~Hammer and M.~J.~Ramsey-Musolf,
  Phys.\ Rev.\ C {\bf 60}, 045204 (1999)
  [Erratum-ibid.\ C {\bf 62}, 049902 (2000)]
  [hep-ph/9903367].

\bibitem{Hammer:1998rz} 
  H.-W.~Hammer and M.~J.~Ramsey-Musolf,
  Phys.\ Rev.\ C {\bf 60}, 045205 (1999)
  [Erratum-ibid.\ C {\bf 62}, 049903 (2000)]
  [hep-ph/9812261].


\bibitem{Sabba Stefanescu:1978sy} 
  I.~Sabba Stefanescu,
  J.\ Math.\ Phys.\  {\bf 21}, 175 (1980).

\bibitem{Sick:2003gm} 
  I.~Sick,
  Phys.\ Lett.\ B {\bf 576}, 62 (2003)
  [nucl-ex/0310008].

%
\bibitem{Arrington:2011kv} 
  J.~Arrington,
  Phys.\ Rev.\ Lett.\  {\bf 107}, 119101 (2011)
  [arXiv:1108.3058 [nucl-ex]].

\bibitem{Blunden:2005jv} 
  P.~G.~Blunden and I.~Sick,
  Phys.\ Rev.\ C {\bf 72}, 057601 (2005)
  [nucl-th/0508037].

\bibitem{Belushkin:2007zv} 
  M.~A.~Belushkin, H.-W.~Hammer and U.-G.~Mei{\ss}ner,
  Phys.\ Lett.\ B {\bf 658}, 138 (2008)
  [arXiv:0705.3385 [hep-ph]].

\bibitem{Arrington:2012dq} 
  J.~Arrington,
  arXiv:1210.2677 [nucl-ex].

\bibitem{Ditsche:2012fv} 
  C.~Ditsche, M.~Hoferichter, B.~Kubis and U.-G.~Mei\ss ner,
  JHEP {\bf 1206}, 043 (2012) [arXiv:1203.4758 [hep-ph]].

\bibitem{Belushkin:2005ds} 
  M.~A.~Belushkin, H.-W.~Hammer and U.-G.~Mei\ss ner,
  Phys.\ Lett.\ B {\bf 633}, 507 (2006)
  [hep-ph/0510382].

\bibitem{Ambrosino:2010bv} 
  F.~Ambrosino {\it et al.}  [KLOE Collaboration],
  Phys.\ Lett.\ B {\bf 700}, 102 (2011)
  [arXiv:1006.5313 [hep-ex]].

\bibitem{Aubert:2009ad} 
  B.~Aubert {\it et al.}  [BABAR Collaboration],
  Phys.\ Rev.\ Lett.\  {\bf 103}, 231801 (2009)
  [arXiv:0908.3589 [hep-ex]].


\bibitem{Bernauer10}
 J. C. Bernauer, 
PhD thesis, Johannes Gutenberg-Universit{\"a}t Mainz, 2010.


\bibitem{Hohler:1976ax} 
  G.~H\"ohler, E.~Pietarinen, I.~Sabba Stefanescu, F.~Borkowski, G.~G.~Simon, V.~H.~Walther and R.~D.~Wendling,
  Nucl.\ Phys.\ B {\bf 114}, 505 (1976).

\bibitem{Mergell:1995bf} 
  P.~Mergell, U.-G.~Mei\ss ner and D.~Drechsel,
  Nucl.\ Phys.\ A {\bf 596}, 367 (1996)
  [hep-ph/9506375].

\bibitem{Hammer:2003ai} 
  H.-W.~Hammer and U.-G.~Mei\ss ner,
  Eur.\ Phys.\ J.\ A {\bf 20}, 469 (2004)
  [hep-ph/0312081].


\bibitem{Riordan:2011}
  S.~Riordan {\it et al.},
  Phys.\ Rev.\ Lett.\  {\bf 105}, 262302 (2010)
  [arXiv:1008.1738 [nucl-ex]].

\bibitem{prep}
I.~T.~Lorenz, H.-W.~Hammer and U.-G.~Mei{\ss}ner, {\it in preparation}.


\end{thebibliography}
\end{document}